\documentclass[aps,
groupedaddress,
nofootinbib,twocolumn]{revtex4-2}

\usepackage[usenames,dvipsnames]{color}
\usepackage{amsfonts,amssymb,amsmath,amsthm}
\usepackage{bm}
\usepackage[a4paper]{geometry}
\usepackage{graphicx}
\usepackage{mathrsfs}
\usepackage[hidelinks]{hyperref}

\newcommand{\mat}[1]{\mathsf{#1}}

\begin{document}

\title{The role of shape operator in gauge theories}

\author{V\'{a}clav Zatloukal}
\email{zatlovac@gmail.com}
\author{\v{S}imon Vedl}

\affiliation{Department of Physics, Faculty of Nuclear Sciences and Physical Engineering,\\
Czech Technical University in Prague, B\v{r}ehov\'{a} 7, 115 19 Praha 1, Czech Republic}

\begin{abstract}
We introduce the concept of shape operator and rotating blade (also known in the theory of embedded Riemannian manifolds as the second fundamental form and the Gauss map) in the realm of Yang-Mills theories. Hence we arrive at new gauge-invariant variables, which can serve as an alternative to the usual gauge potentials.


\end{abstract}

\maketitle

\section{Introduction}

Gauge symmetries play a fundamental role in the current formulation of physics as they provide a unifying principle from which all fundamental interactions of the standard model, as well as gravity, derive \cite{Ramond,Baez}.

This \emph{principle of gauge invariance}, i.e., invariance of the theory under local transformations, leads to the introduction of covariant derivatives of the form \cite{YangMills1954}
\begin{equation}
D_\mu \psi
= \partial_\mu \psi + i \mat{A}_\mu \psi .
\label{CovDerYM}
\end{equation}
Throughout this paper we omit the coupling constant, and will assume, for definiteness, that the (matter) field $\psi$ takes values in $\mathbb{C}^n$, and the gauge group is the unitary group $U(n)$ (or its subgroup). The gauge potential ${\mat{A} = \mat{A}_\mu dx^\mu}$ is then a differential 1-form with values in $n$ by $n$ hermitian matrices, i.e., $i \mat{A}_\mu(x)$ are elements of the Lie algebra $\mathfrak{u}(n)$.
In particular, for the Abelian case of gauge group $U(1)$, the $A_\mu$ are components of the electromagnetic four-potential. 

The field intensities (or field strengths) are encapsulated in the curvature two-form ${\mat{F} = d\mat{A} + i\mat{A} \wedge \mat{A}}$, whose components are given by the commutator of covariant derivatives:
\begin{align}
\mat{F}_{\mu\nu}
&= -i [D_\mu,D_\nu]
\nonumber\\
&= \partial_\mu \mat{A}_\nu - \partial_\nu \mat{A}_\mu 
+ i[\mat{A}_\mu,\mat{A}_\nu] .
\label{FieldStrYM}
\end{align}
In the case of electromagnetism it reduces to the Faraday tensor ${F_{\mu\nu} = \partial_\mu A_\nu - \partial_\nu A_\mu}$, which assembles both electric and magnetic intensities.

Although Maxwell's equations of classical electromagnetism are formulated purely in terms of the field intensities, the four-potential is a practical tool in their analysis. More importantly, it allows one to develop the action principle \cite[Ch.\,4]{LandauLifshitz}, and lay electromagnetism within quantum theory. At the same time, however, the gauge freedom involves redundancy in the description of physical configurations, which needs to be dealt with by gauge fixing, or, as in the Faddeev-Popov procedure of the path-integral formulation \cite[Ch.\,9.4]{PeskinSchroeder}, by factoring out the volume of the gauge orbit.

The purpose of this article is to present new variables -- the \emph{rotating blade}, and its derivative, the \emph{shape operator} -- which provide an alternative description of electromagnetism as well as general Yang-Mills theories. These quantities are rather standard in the context of embedded Riemannian manifolds, where the rotating blade represents the moving tangent plane, and the shape operator is a measure of extrinsic curvature \cite[Ch.\,6.5]{DoranLasenby}\cite{Hestenes2011}. We review these facts about embedded manifolds in Section~\ref{sec:EmbMan}, and in Section~\ref{sec:ShapeOpYM} introduce analogous quantities in the realm of Yang-Mills theories.

An essential step, as we shall see, is to solve the equation
\begin{equation}
\mat{V}^\dag \partial_\mu \mat{V}
= i\mat{A}_\mu ,
\label{Vequation}
\end{equation}
where $\mat{V}(x)$ takes values in complex matrices with $N \geq n$ rows and $n$ columns satisfying the constraint ${\mat{V}^\dag \mat{V} = \mat{I}}$. (By $\mat{I}$ we denote the identity matrix with appropriate dimension.) For large enough $N$, Eq.~\eqref{Vequation} can be solved for any gauge potential $\mat{A}_\mu$. This result of the theory of \emph{universal connections}\footnote{The theory asserts that for a principal bundle over an arbitrary base manifold, any principal connection can be obtained via pullback of a universal (canonical) connection. The latter is defined on a principal bundle over the manifold of $n$-dimensional subspaces of an $N$-dimensional linear space (the Grassmannian).} holds for arbitrary compact gauge group \cite{Narasimhan1961} (in fact, even for arbitrary connected Lie group \cite{Narasimhan1963}).  In Appendix \ref{sec:RotBladeExistence} a simple proof in the case of electromagnetism is provided, which gives $N = 4$. 

Having the function $\mat{V}$ at our disposal, we introduce the rotating blade to be the manifestly gauge-invariant quantity ${\mat{R} = 2 \mat{V} \mat{V}^\dag - \mat{I}}$, from which the shape operator is obtained by differentiation: $\mat{S}_\mu = -\frac{i}{2} \mat{R} \partial_\mu \mat{R}$.

Further in Section~\ref{sec:ShapeOpYM} we point out that the shape operator can be viewed as a special gauge, the \emph{shape gauge}, derived from $\mat{A}_\mu$ via a gauge transformation within the extended gauge group $U(N)$. Also, we briefly discuss dynamics, and show that the standard Yang-Mills action yields less restrictive equations of motion when formulated in terms of the variable $\mat{V}$.

In Section~\ref{sec:Examples} we work in the simplest scenario $n=1$ and $N=2$, and presents two examples of electromagnetic fields cast in the rotating blade representation: planar wave and magnetic monopole. Although our investigations in this article are mainly local, the magnetic monopole example suggests that the rotating blade variable could be rather useful when it comes to global aspects of gauge theories.

\section{Shape operator for embedded manifolds}
\label{sec:EmbMan}

We consider a $d$-dimensional manifold $\mathcal{M}$, with coordinates $x = (x^\mu)$, embedded in $\mathbb{R}^N$ (endowed with standard scalar product `$\,\cdot\,$') by means of a smooth function ${f: \mathcal{M} \rightarrow \mathbb{R}^N}$. 
The tangent space is spanned by the vectors ${f_\mu \equiv \partial_\mu f}$ whose scalar product induces a metric ${g_{\mu\nu} = f_\mu \cdot f_\nu}$ on $\mathcal{M}$. 

The covariant derivative of a tangent vector field $v$ is commonly defined \cite[Ch.\,9.1]{Frankel} by projecting (by the projector $\mat{P}$) the result of partial differentiation back onto the tangent space: $\mathcal{D}_\mu v = \mat{P} \partial_\mu v$. Denoting by ${\mat{P}_\perp = \mat{I} - \mat{P}}$ the projection on the normal space (the orthogonal complement of the tangent space), the covariant derivative of any vector field with values in $\mathbb{R}^N$ is defined as
\begin{align}
\mathcal{D}_\mu v 
&= \mat{P}\partial_\mu(\mat{P} v) + \mat{P}_\perp\partial_\mu(\mat{P}_\perp v)
\nonumber\\
&= \partial_\mu v + \mat{S}_\mu v,
\label{eq:EmbeddedCovariantDerivative}
\end{align}
where we have introduced the \emph{shape operator}\footnote{In classical differential geometry, the shape operator (or Weingarten map) of a codimension-1 hypersurface is introduced via differentiation of the unit normal field $n$ \cite[Ch.\,5]{ONeil}\cite[Ch.\,13]{Gray}. This gives the components of matrices $\mat{S}_\mu$ in Eq.~\eqref{ShapeEmb}, since, by a simple calculation, $$f_\nu \cdot \mat{S}_\mu (n) = -f_\nu \cdot (\partial_\mu n) = (\partial_\mu f_\nu) \cdot n .$$ Note also that the right-hand side are the coefficients of the second fundamental form \cite[p.\,33]{Spivak3}.}
\begin{align}
\mat{S}_\mu 
&= \mat{P}\partial_\mu \mat{P} + \mat{P}_\perp\partial_\mu\mat{P}_\perp
\nonumber\\
&= \frac{1}{2}\mat{R}\partial_\mu\mat{R} .
\label{ShapeEmb}
\end{align}
Here $\mat{R} = 2\mat{P} - \mat{I}$ is reflection with respect to the tangent plane. Since $\mat{R}^T = \mat{R}$, and $\mat{R}^2 = \mat{I}$, $\mat{S}_\mu$ is for each $\mu$ a skew-symmetric linear operator that maps normal vectors to the tangent space and vice versa.

\begin{figure}[h]
\includegraphics[scale=0.85]{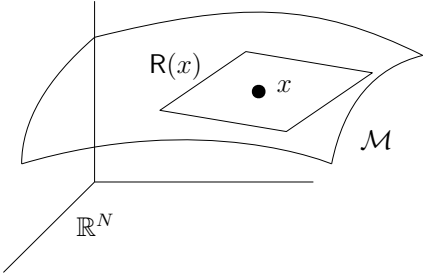}
\caption{A manifold $\mathcal{M}$ embedded in $\mathbb{R}^N$. The tangent space at point $x$ is represented by a reflection matrix $\mat{R}(x)$ (the rotating blade).}
\label{fig:Riem}
\end{figure}

For each point of the manifold, the tangent space can be identified with a $d$-dimensional linear subspace of $\mathbb{R}^N$, and hence is an element of the (real) Grassmannian manifold $Gr_{\mathbb{R}}(d,N)$. We represent tangent spaces explicitly by reflections $\mat{R}$ (or, equivalently, projections $\mat{P}$), and call the function $\mat{R}(x)$ (or $\mat{P}(x)$) the \emph{rotating blade}.\footnote{The term `blade' is used in geometric algebra for decomposable multivectors of a real Clifford algebra \cite{HestenesSobczyk}. These provide yet another representation of linear subspaces.} (See Fig.~\ref{fig:Riem}.) In more abstract terms, rotating blade is the generalized Gauss map \cite[Ch.\,VII-2]{KobNom}, a mapping from $\mathcal{M}$ to $Gr_{\mathbb{R}}(d,N)$.

In view of Eq.~\eqref{ShapeEmb} the rotating blade can be thought of as a ``potential" for the shape operator, which then satisfies the identity
\begin{equation}
\partial_\mu \mat{S}_\nu - \partial_\nu \mat{S}_\mu 
= -2[\mat{S}_\mu,\mat{S}_\nu].
\label{eq:EmbeddedShapeEquation}
\end{equation}
This observation can be used to simplify the curvature $\mat{\Omega}_{\mu\nu} = [\mathcal{D}_\mu, \mathcal{D}_\nu]$ to 
\begin{align}
\mat{\Omega}_{\mu\nu}
&= -[\mat{S}_\mu,\mat{S}_\nu]
\nonumber\\
&= \frac{1}{4}[\partial_\mu\mat{R},\partial_\nu\mat{R}] .
\label{eq:EmbeddedCurvatureCommutator}
\end{align}
That is, the curvature $\mat{\Omega}_{\mu\nu}$ is determined algebraically from the shape operator, or from first derivatives of the rotating blade. 

At this point it is worth to recall that to calculate the intrinsic (Riemann) curvature of an abstract Riemannian manifold one needs the second derivatives of the metric tensor $g_{\mu\nu}$ (or the first derivatives of the connection coefficients -- the Christoffel symbols).  Meanwhile, in the embedded case the intrinsic curvature can be obtained from the tangent part of Eq.~\eqref{eq:EmbeddedCurvatureCommutator}: ${R_{\rho\sigma\mu\nu} = f_\rho\cdot(\mat{\Omega}_{\mu\nu}f_\sigma})$. There is, therefore, a certain trade-off between the differential complexity (the degree of derivatives needed) in the intrinsic approach, and the algebraic complexity (the number of extra dimensions of the ambient space) in the embedded approach.


\section{Shape operator and rotating blade for Yang-Mills theories}
\label{sec:ShapeOpYM}

We now wish to define appropriate analogues of the shape operator and the rotating blade in the realm of $U(n)$ Yang-Mills theories. These objects will be denoted by the same symbols as in the previous section, although the respective equations will differ by some conventional extra factors of $i$ (the imaginary unit).

We start with a $d$-dimensional spacetime $\mathcal{M}$, an $n$-component field $\psi$, and a covariant derivative given by Eq.~\eqref{CovDerYM}.
Finding an $N \times n$-matrix-valued function $\mat{V}$ that satisfies Eq.~\eqref{Vequation}, the covariant derivative acquires the form
\begin{align}
D_\mu \psi
&= \partial_\mu \psi + i \mat{A}_\mu \psi
\nonumber\\
&= \mat{V}^\dag \partial_\mu (\mat{V} \psi) .
\label{CovDerV}
\end{align}
At each point of the spacetime the columns of $\mat{V}$ define an $n$-dimensional linear subspace of $\mathbb{C}^N$ along with a choice of its orthonormal basis.
The covariant derivative as expressed in Eq.~\eqref{CovDerV} has appealing geometric interpretation: $\psi$ is being lifted to the subspace defined by $\mat{V}$, then differentiated by means of a flat partial derivative, and finally projected by $\mat{V}^\dag$ back to its original $\mathbb{C}^n$ space.

\begin{figure}[h]
\includegraphics[scale=0.85]{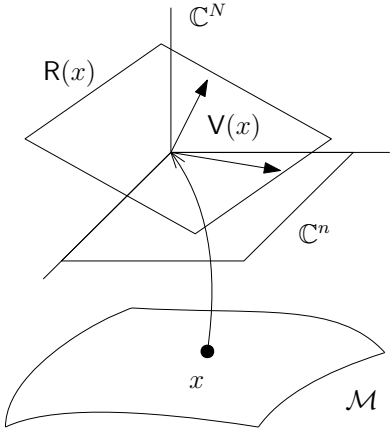}
\caption{For a $U(n)$ Yang-Mills theory, the rotating blade $\mat{R}(x)$ defines at each point of the spacetime manifold $\mathcal{M}$ an $n$-dimensional linear subspace of the extended internal space $\mathbb{C}^N$. This subspace is spanned by the columns of the matrix $\mat{V}(x)$.}
\label{fig:YM}
\end{figure}

Having subspaces defined by the function $\mat{V}$ (or correspondingly by the projector $\mat{P} = \mat{V}\mat{V}^\dag$) at our disposal, we can introduce, in complete analogy with Eq.~\eqref{eq:EmbeddedCovariantDerivative}, a covariant derivative that acts on any $\mathbb{C}^N$-valued field $\Psi$,
\begin{align}
\mathcal{D}_\mu \Psi 
&= \mat{P}\partial_\mu(\mat{P} \Psi) + \mat{P}_\perp\partial_\mu(\mat{P}_\perp \Psi)
\nonumber\\
&= \partial_\mu \Psi + i \mat{S}_\mu \Psi,
\label{CovDerProjYM}
\end{align}
and which for $\Psi = \mat{V}\psi$ reduces to $\mat{V}D_\mu \psi$.\footnote{Note that in comparison with Eq.~\eqref{ShapeEmb} we added a factor $i$ in front of the shape operator $\mat{S}_\mu$ to parallel the expression~\eqref{CovDerYM} of Yang-Mills covariant derivative.}
The Yang-Mills shape operator reads
\begin{align}
\mat{S}_\mu 
&= -\frac{i}{2}\mat{R}\partial_\mu\mat{R} ,
\label{ShapeYM}
\end{align}
where again $\mat{R} = 2 \mat{P} - \mat{I}$ is the reflection with respect to the subspace defined by $\mat{V}$, and the function $\mat{R}(x)$ will be referred to as the (Yang-Mills) rotating blade\footnote{We shall use this denomination with either the abstract subspace or its representation by the projector $\mat{P}$ or the reflection $\mat{R}$ in mind as each one of these defines uniquely and straightforwardly the others.} (see Fig.~\ref{fig:YM}). Note that since ${\mat{R} \,\mat{S}_\mu = - \mat{S}_\mu \mat{R}}$, the rotating blade is covariantly constant: ${\mathcal{D}_\mu \mat{R} = 0}$.

The curvature of $\mathcal{D}_\mu$ is given by (cf. Eq.~\eqref{FieldStrYM})
\begin{align}
\mat{\Omega}_{\mu\nu}
&= -i [\mathcal{D}_\mu, \mathcal{D}_\nu]
\nonumber\\
&= -i [\mat{S}_\mu, \mat{S}_\nu]
\nonumber\\
&= -\frac{i}{4} [\partial_\mu \mat{R}, \partial_\nu \mat{R}]
\nonumber\\
&= -i [\partial_\mu \mat{P}, \partial_\nu \mat{P}] ,
\label{CurvOmega}
\end{align}
where we have made use of the fact that ${\mat{R}^2 = \mat{I}}$, and hence
\begin{equation}
\partial_\mu \mat{S}_\nu - \partial_\nu \mat{S}_\mu 
= -2i[\mat{S}_\mu,\mat{S}_\nu].
\end{equation}
We remark that the latter equation can be neatly written as
\begin{equation}
\mathcal{D}_\mu \mat{S}_\nu 
= \mathcal{D}_\nu \mat{S}_\mu .
\end{equation}

A change of gauge implemented by a $U(n)$-valued function $\mat{u}(x)$ entails the transformations
\begin{align}
\psi' 
&= \mat{u}\psi
\nonumber\\
\mat{A}'_\mu 
&= \mat{u} \mat{A}_\mu \mat{u}^\dag - i \mat{u} \partial_\mu \mat{u}^\dag
\nonumber\\
\mat{F}'_{\mu\nu}
&= \mat{u} \mat{F}_{\mu\nu} \mat{u}^\dag
\nonumber\\
\mat{V}'
&= \mat{V} \mat{u}^\dag ,
\end{align}
where the first three rules are standard, and the last one has been postulated in accordance with Eq.~\eqref{Vequation}. The projection $\mat{P}$, as well as the derived quantities $\mat{R}$, $\mat{S}_\mu$, and $\mat{\Omega}_{\mu\nu}$, are then manifestly invariant under $U(n)$ gauge transformations, and the same holds for the product $\mat{V}\psi$. Therefore, by  lifting the `matter' field $\psi$ to $\Psi = \mat{V}\psi$, and the covariant derivative $D_\mu$ to $\mathcal{D}_\mu$, the $U(n)$ gauge has been eliminated from the formalism, and it only reemerges once we choose an orthonormal frame for the rotating blade, i.e. assign a concrete $\mat{V}$ to the projection $\mat{P}$.

There is, in fact, a preferred way how to make this assignment, and hence choose a gauge for $\mat{A}_\mu$.
To see this, we write ${\mat{V} = \mat{U} \mat{V}_0}$, where ${\mat{V}_0 = (\mat{I}, \mat{0})^T}$, and ${\mat{U}(x) \in U(N)}$. Next, we make use of the Cartan decomposition \cite[Ch.\,VI.3]{Knapp}, which implies that any unitary matrix can be uniquely decomposed
into a product of unitaries ${\mat{U} = \mat{U}_1 \mat{U}_2}$, such that ${\mat{U}_2 \mat{R}_0 = \mat{R}_0 \mat{U}_2}$, while ${\mat{U}_1 \mat{R}_0 = \mat{R}_0 \mat{U}^\dag_1}$. (Here $\mat{R}_0$ is the reflection corresponding to $\mat{V}_0$.) 
It follows that
\begin{equation}
\mat{V}
= \mat{U}_1 \mat{U}_2 \mat{V}_0
= \mat{U}_1 \mat{V}_0 \mat{U}'_2 ,
\end{equation}
where $\mat{U}'_2 = \mat{V}_0^\dag \mat{U}_2 \mat{V}_0$ is $n$ by $n$ unitary. The choice of $\mat{U}_2$ has no effect on the rotating blade $\mat{R}$, and we may decide to fix it as $\mat{U}_2 = \mat{I}$. This yields the gauge-fixed potential
\begin{equation}
i\mat{A}_\mu 
= \mat{V}_0^\dag (\mat{U}_1^\dag \partial_\mu \mat{U}_1) \mat{V}_0 .
\end{equation}

\subsection{Shape gauge}
\label{sec:ShapeGauge}

To better understand the relationship between the covariant derivatives $D_\mu$ and $\mathcal{D}_\mu$ we extend $\mat{V}$ at each point to an $N$ by $N$ unitary matrix $\mat{U} = (\mat{V}, \mat{W})$, i.e., choose (at each point) an orthonormal basis of the orthogonal complement of the rotating blade. The shape operator can then be viewed as a result of a $U(N)$ gauge transformation, 
\begin{equation}
\mat{S}_\mu 
= \mat{U}
\begin{pmatrix}
\mat{A}_\mu & \mat{0} \\
\mat{0} & \mat{C}_\mu
\end{pmatrix}
 \mat{U}^\dag - i \mat{U} \partial_\mu \mat{U}^\dag ,
\end{equation}
of a combined gauge potential $\mat{A}_\mu \oplus \mat{C}_\mu$ acting on fields with values in $\mathbb{C}^n \oplus \mathbb{C}^{N-n}$, where ${i\mat{C}_\mu = \mat{W}^\dag \partial_\mu \mat{W}}$.

The corresponding curvatures are gauge-related:
\begin{equation}
\mat{\Omega}_{\mu\nu}
= \mat{U} 
\begin{pmatrix}
\mat{F}_{\mu\nu} & \mat{0} \\
\mat{0} & \mat{G}_{\mu\nu}
\end{pmatrix}
 \mat{U}^\dag ,
\label{Omega}
\end{equation}
where $\mat{G}_{\mu\nu}$ denotes the curvature of the complementary connection $\mat{C}_\mu$. Conversely, we can write ${\mat{F}_{\mu\nu} = \mat{V}^\dag \mat{\Omega}_{\mu\nu} \mat{V}}$, and ${\mat{G}_{\mu\nu} = \mat{W}^\dag \mat{\Omega}_{\mu\nu} \mat{W}}$.

It is important to distinguish between the `small' $U(n)$ gauge transformations $\mat{u}$, under which $\mat{S}_\mu$ is invariant, and `big' $U(N)$ gauge transformations $\mat{U} = (\mat{V}, \mat{W})$, which relate ${\mat{A}_\mu \oplus \mat{C}_\mu}$ with a \emph{shape gauge} $\mat{S}_\mu$.

In this context it should be noted that the shape gauge is not unique, i.e., there can be various $\mat{S}_\mu$ corresponding to a given gauge potential $\mat{A}_\mu$, as there can exist various solutions $\mat{V}$ of Eq.~\eqref{Vequation} (from which the shape operator derives). Namely, from any solution $\mat{V}$ we obtain another solution $\mat{U}_1 \mat{V}$ whenever the $U(N)$-valued field $\mat{U}_1$ satisfies $\mat{V}^\dag (\mat{U}_1^\dag \partial_\mu \mat{U}_1) \mat{V} = \mat{0} $.

\subsection{Rotating blade dynamics}

For flat spacetime, and no currents, the dynamics of the gauge fields $\mat{A}_\mu$, and their intensities $\mat{F}_{\mu\nu}$ is controlled by the Yang-Mills equations \cite[Ch.\,6.3]{Ramond}
\begin{equation}
D^\mu \mat{F}_{\mu\nu} 
= \mat{0} ,
\label{YMeq}
\end{equation}
which reduce to vacuum Maxwell's equations ${\partial^\mu F_{\mu\nu} 
= 0}$ in the case of electromagnetism.\footnote{The covariant derivative acts on matrix-valued fields (with gauge transformation ${\mat{M}' = \mat{u} \mat{M} \mat{u}^\dag}$) as
\begin{equation*}
D_\mu \mat{M}
= \partial_\mu \mat{M} + i [ \mat{A}_\mu, \mat{M}] .
\end{equation*}}

Eq.~\eqref{YMeq} can also be derived by varying the action
\begin{equation}
\mathcal{S}_{YM}[\mat{A}_\mu]
= -\frac{1}{4}\int d^d x\operatorname{Tr}(\mat{F}_{\mu\nu}\mat{F}^{\mu\nu}) .
\label{YMaction}
\end{equation}
It is worth to note that expressing $\mat{A}_\mu$ in terms of $\mat{V}$, through Eq.~\eqref{Vequation}, the equations that result from variation of the action $\mathcal{S}_{YM}[\mat{V}]$ are less restrictive (i.e., they allow more solutions) than the Yang-Mills equations~\eqref{YMeq} (regarded as equations for $\mat{V}$). 

To see this, we parametrize infinitesimal variations that respect the constraint $\mat{V}^\dag \mat{V} = \mat{I}$ as $\mat{V} \rightarrow e^{i \delta\mat{B}} \mat{V}$, with $\delta\mat{B}^\dag = \delta\mat{B}$, to find
\begin{equation}
\delta \mat{A}_\mu 
= \mat{V}^\dag ( \partial_\mu \delta\mat{B} ) \mat{V} ,
\end{equation}
and hence the equation of motion
\begin{equation}
\partial^\nu \big(\mat{V} (D^\mu \mat{F}_{\mu\nu}) \mat{V}^\dag \big)
= \mat{0} .
\end{equation}
For electromagnetism this reduces to
\begin{equation}
(\partial^\mu F_{\mu\nu})\, \partial^\nu \mat{R}
= \mat{0} .
\label{MaxwMod}
\end{equation}

Note that the Yang-Mills equations~\eqref{YMeq} can be cast as equations for the rotating blade when transformed to the shape gauge. We obtain
\begin{equation}
\mat{P}\, \mathcal{D}^\mu \mat{\Omega}_{\mu\nu} 
= \mat{0} ,
\label{YMshape}
\end{equation}
where the projection $\mat{P}$ picks out only the $\mat{F}$-part of Eq.~\eqref{Omega}. 



Let us make one more comment. The Yang-Mills Lagrangian in \eqref{YMaction} is often described as the simplest gauge-invariant Lorentz-scalar quantity one can construct out of the degrees of freedom $\mat{A}_\mu$ \cite[p.196]{Ramond}. In fact, for the rotating blade $\mat{R}$ one can consider an even simpler action
\begin{equation}
\mathcal{S}_\sigma[\mat{R}]
= -\frac{1}{4}\int d^d x \operatorname{Tr}(\partial_\mu\mat{R}\,\partial^\mu\mat{R}) ,
\label{eq:SigmaModelAction}
\end{equation}
defining a nonlinear Grassmannian $\sigma$-model \cite{Brezin,Holten,Pisarski}.
The equation of motion
\begin{equation}
\partial_\mu \mat{S}^\mu 
= \mat{0} ,
\label{eq:SigmaModelEOM}
\end{equation}
which follows from variations $\mat{R} \rightarrow e^{i \delta\mat{B}} \mat{R} e^{-i \delta\mat{B}}$, is trim, but its physical relevance, i.e., implications for the field intensities $\mat{F}_{\mu\nu}$, is not clear.

\section{Electromagnetism -- examples}
\label{sec:Examples}

We will now investigate the $U(n=1)$ gauge theory -- the electromagnetism -- and present two concrete examples of rotating blades for $N=2$. 

The matrix $\mat{V}$ then reduces to a normalized column vector
\begin{equation}
\mat{V}
= \begin{pmatrix} e^{i\alpha} \cos\rho
\\ e^{i\beta} \sin\rho \end{pmatrix} ,
\label{Vem}
\end{equation}
parametrized by three real-valued functions $\alpha$, $\beta$ and $\rho$.
The ensuing rotating blade
\begin{equation}
\mat{R}=\begin{pmatrix}
\cos 2\rho & e^{i(\alpha-\beta)}\sin 2\rho  \\
e^{-i(\alpha-\beta)}\sin 2\rho  & -\cos 2\rho
\end{pmatrix} 
\label{RotBladeEM}
\end{equation}
has one degree of freedom fewer, as it is invariant under a common shift of $\alpha$ and $\beta$ (which corresponds to a $U(1)$ gauge transformation).
Eq.~\eqref{Vequation} becomes
\begin{equation}
\cos^2\!\rho \,\partial_\mu\alpha
+ \sin^2\!\rho \,\partial_\mu\beta
= A_\mu ,
\label{VeqEM}
\end{equation}
or $\cos^2\!\rho \,d\alpha
+ \sin^2\!\rho \,d\beta = A$ in the language of differential forms.

The Faraday tensor, i.e., the curvature of $D_\mu$, is given by the exterior differential
\begin{equation}
F = dA = d(\cos^2\rho) \wedge d(\alpha-\beta) .
\end{equation}
This is a decomposable 2-form, which always satisfies $F \wedge
F = 0$, and hence the condition $\textbf{B} \cdot \textbf{E} = 0$ between the electric and the magnetic field \cite[Sec.\,25]{LandauLifshitz}. This condition can be removed by taking $N=4$ (see Appendix~\ref{sec:RotBladeExistence}).

Let us also remark that the vector  
\begin{equation}
\mat{W}
= \begin{pmatrix} -e^{-i\beta} \sin\rho
\\ e^{-i\alpha} \cos\rho \end{pmatrix} ,
\end{equation}
is an orthogonal complement of $\mat{V}$, which yields the complementary connection $C_\mu = -A_\mu$, and its field strength $G_{\mu\nu} = -F_{\mu\nu}$.

\subsection{Example: Plane wave}

The electromagnetic plane wave with wave-vector $k_\mu$ and polarization $n_\mu$ is characterized by the four-potential
\begin{equation}
A_\mu = n_\mu \sin(k_\nu x^\nu) ,
\end{equation}
and Eq.~\eqref{VeqEM} is solved by the functions
\begin{align}
\alpha &= n_\mu x^\mu
\nonumber\\
\beta &= -n_\mu x^\mu
\nonumber\\
\rho &= \frac{1}{2} k_\mu x^\mu - \frac{\pi}{4} .
\end{align}

Looking back at Eq.~\eqref{MaxwMod} we note that its plane-wave solutions must satisfy the condition ${(k_\mu k^\mu)(n_\nu n^\nu) = (k_\mu n^\mu)^2}$, which features the four-vectors $k$ and $n$ completely symmetrically. This condition is weaker than the conditions on plane-wave solutions of Maxwell's equations, $k_\mu k^\mu = k_\mu n^\mu = 0$.

\subsection{Example: Magnetic monopole}
\label{sec:Monopole}

Next we consider a magnetic monopole with strength $g$, i.e., a radial magnetic field configuration  
\begin{equation}
\textbf{B} = g \frac{\textbf{x}}{r^3} .
\label{MonB}
\end{equation}
In spherical coordinates $(r,\theta,\varphi)$ this can be described by a pair\footnote{If a single global potential $A$ existed, then the integral of $F=dA$ over a sphere centered at the origin would have to vanish as a result of the Stokes theorem, but at the same time be equal to $4\pi g$ due to the radial form of the magnetic field~\eqref{MonB}.} of potentials \cite[Ch.\,16.4e]{Frankel}
\begin{equation}
A^{(\pm)}
= g(\pm 1 - \cos\theta) d\varphi ,
\end{equation}
where $A^{(+)}$ is not defined for $\theta = \pi$, while $A^{(-)}$ is not defined for $\theta = 0$. On the overlap of their domains, the potentials are gauge-related: ${A^{(+)} = A^{(-)} + 2g d \varphi}$. 

Eq.~\eqref{VeqEM} is solved by
\begin{align}
\alpha^{(+)} &= 0 ~,~ \alpha^{(-)} = -2 g \varphi
\nonumber\\
\beta^{(+)} &= 2 g \varphi ~,~ \beta^{(-)} = 0
\nonumber\\
\rho^{(\pm)} &= \frac{\theta}{2} ,
\end{align}
which, when plugged into Eq.~\eqref{RotBladeEM} yields a single rotating blade ${\mat{R} = \mat{R}^{(+)} = \mat{R}^{(-)}}$. If the (quantization) condition $2 g \in \mathbb{Z}$ is met, this $\mat{R}$ is defined on the whole of $\mathbb{R}^3$ except at the origin.

Let us remark that a similar description of magnetic monopole was achieved in Ref.~\cite{Ryder1980}, with our quantity $\mat{V}$ in Eq.~\eqref{Vem} regarded as an element of a $3$-sphere $S^3$.



\section{Conclusion}

We introduced the rotating blade variable, and its derivative -- the shape operator -- in the realm of Yang-Mills theories, borrowing intuition from analyses of embedded manifolds. 
\footnote{We found particularly inspiring the ideas of Refs.~\cite{HestenesSobczyk,Hestenes2011,DoranLasenby}, where the fundamental variable describing the geometry of a manifold is its \emph{pseudoscalar} field, i.e., a multivector representation of the tangent space. In the context of gauge theories we prefer to use a different term -- the rotating blade -- since we are putting emphasis on the fibre bundle connection rather than on the underlying spacetime manifold.}
In view of Eq.~\eqref{CovDerProjYM} the shape operator plays the role of connection coefficients of the covariant derivative $\mathcal{D}_\mu$, and at the same time `factorizes' the curvature $\Omega_{\mu\nu}$ in the sense of Eq.~\eqref{CurvOmega}.

The essential technical observation in our approach was that via Eq.~\eqref{Vequation} any $U(n)$ gauge potential $\mat{A}_\mu$ can be represented by sufficiently large matrix $\mat{V}$, which defines an $n$-tuple of orthogonal vectors in $\mathbb{C}^N$. Factoring out the internal rotations of this $n$-tuple we arrived at the rotating blade $\mat{R}$, which thus captures $\mat{A}_\mu$ together with all gauge-equivalent potentials, i.e., the entire gauge orbit. In the rotating-blade formulation, the Yang-Mills theories are no longer \emph{gauge} theories, since there is no gauge redundancy any more. At the same time, however, it should be stressed that the gauge orbit is not represented by a unique rotating blade, so it looks as if the gauge ambiguity has been replaced by another type of ambiguity (see the comment at the end of Sec.~\ref{sec:ShapeGauge}). 

It remains to be seen whether rotating blades could present a viable alternative to the traditional gauge potentials. 
Let us consider, for example, the Yang-Mills equations. In their traditional form, Eq.~\eqref{YMeq}, these are second-order partial differential equations for the gauge potential $\mat{A}_\mu$, which simplify to linear (Maxwell's) equations in the case of electromagnetism. Their shape-gauge image, Eq.~\eqref{YMshape}, is a second-order equation for $\mat{R}$ (first-order for the shape operator $\mat{S}_\mu$), which is nonlinear also for the electromagnetic theory. 
One can thus expect certain technical challenges when working with rotating blades.

Let us conclude with an intriguing 
quote of David Hestenes \cite[p.\,11]{Hestenes2011} presented during his analysis of manifolds: ``\ldots the treatment of intrinsic geometry can be simplified by coordinating it with extrinsic geometry!". Our aim has been to suggest that this could be the case, in some respect, also in the context of gauge theories and connections on fibre bundles.

\begin{acknowledgments}
We would like to thank Alexander Thomas for valuable discussions. \v{S}.V. was supported by the Grant Agency of the Czech Technical University in Prague, grant No. SGS22/178/OHK4/3T/14.
\end{acknowledgments}

\appendix
\section{Rotating blade for electromagnetism}
\label{sec:RotBladeExistence}

In the case of electromagnetism, i.e. a ${U(n=1)}$ gauge theory, on a $d$-dimensional spacetime, we take advantage of the fact (the Darboux theorem \cite[p.\,40]{EDS}) that the $1$-form $A=A_\mu dx^\mu$ can be locally expressed as 
\begin{equation} \label{Acan}
A = \sum_{k=0}^r \pi_k d\phi_k ,
\end{equation}
where $\{ \pi_k, \phi_k \}$ are independent functions. Here $r < \frac{d}{2}$ is the \emph{rank} of the differential $1$-form $A$, defined by the conditions
\begin{equation}
A\wedge(dA)^r \neq 0,\quad A\wedge(dA)^{r+1} = 0 .
\end{equation}

The corresponding rotating blade is given by $\mat{R} = 2 \mat{V} \mat{V}^\dag - \mat{I}$, where $\mat{V}$ a solution of Eq.~\eqref{Vequation} constructed as follows:
\begin{equation}
\mat{V}
= \frac{1}{\sqrt{r+1}}
\begin{pmatrix}
\mat{V}_0 \\
\vdots \\
\mat{V}_r
\end{pmatrix} ,
\end{equation}
where the two-component vectors (see Eq.~\eqref{Vem})
\begin{equation}
\mat{V}_k = 
\begin{pmatrix} e^{i\alpha_k}\cos\rho_k \\ e^{i\beta_k}\sin\rho_k \end{pmatrix} \label{eq:TwoCompBuildBlock}
\end{equation}
satisfy
\begin{equation}
\cos^2\!\rho_k\, d\alpha_k + \sin^2\!\rho_k\,d\beta_k
= \pi_k d\phi_k 
\end{equation}
(no sum over $k$). The latter equation is solved, for example, by 
\begin{align}
\alpha_k
&= -\beta_k
= \phi_k ,
\nonumber\\
\rho_k
&= \frac{1}{2} \arccos \pi_k ,
\end{align}
where we have assumed that the functions $\pi_k$ are scaled so that on a local neighbourhood of interest $|\pi_k| \leq 1$.

The vector $\mat{V}$ has $N = 2(r+1)$ components, which gives $N=4$ in the case of four-dimensional spacetime. This is an improvement over the algorithm of Ref.~\cite[Sec.\,3]{Narasimhan1961}, which yields $N=9$. However, the latter algorithm is purely algebraic, whereas our construction requires solving the differential problem of finding the Darboux coordinates $\{\pi_k,\phi_k\}$, to be used in Eq.~\eqref{Acan}.

\end{document}